\def\Bbb#1{{\bf #1}}

\def\fnote#1{\footnote}

\def\cwleftpar#1#2{\leftskip #1 \rightskip #2 plus 1fill}
\def\cwrightpar#1#2{\leftskip #1 plus 1fill \rightskip #2}
\def\cwcenterpar#1#2{\leftskip #1 plus 1fill \rightskip #2 plus 1fill}
\def\cwfullpar#1#2{\leftskip#1\rightskip#2}

\def\cwoutdent#1#2{\llap{\hbox to #1{#2 \hss}}\ignorespaces}
\def\cwparbegin#1#2#3#4#5{
	\ifcase #1 \cwleftpar{#2}{#3}
	\or \cwrightpar{#2}{#3}
	\or \cwcenterpar{#2}{#3}
	\else \cwfullpar{#2}{#3}\fi
	\ifcase #4 \baselineskip = 1.5\baselineskip
	\or \baselineskip = 2\baselineskip
	\or \baselineskip = 3\baselineskip
	\else \baselineskip = 1\baselineskip\fi
	\ifdim #5 > 0in \else \noindent \fi
	\noindent\ignorespaces}
\documentclass{article}
\begin{document}
\advance \vsize by -1\baselineskip
\def\makefootline{
\ifnum\pageno = 1{\vskip \baselineskip \vskip \baselineskip }\else{\vskip \baselineskip \noindent \folio                                  \par
}\fi}

 \vspace*{3ex}
\noindent {\bf JOINT INSTITUTE FOR NUCLEAR RESEARCH}\\[2ex]

\noindent {\bf Bogoliubov Laboratory of Theoretical Physics}

 \vspace*{4ex}

\noindent {\Huge First Order Deviation Equations in\\[1ex]
	 Spaces with a Transport along Paths}

\vspace*{5ex}

\noindent Bozhidar Zakhariev Iliev
\fnote{0}{\noindent $^{\hbox{}}$Permanent address:
Laboratory of Mathematical Modeling in Physics,
Institute for Nuclear Research and \mbox{Nuclear} Energy,
Bulgarian Academy of Sciences,
Boul.\ Tzarigradsko chauss\'ee~72, 1784 Sofia, Bulgaria\\
\indent E-mail address: bozho@inrne.bas.bg\\
\indent URL: http://theo.inrne.bas.bg/$^\sim$bozho/}

\vspace*{12ex}
{\bf \noindent Published: Communication JINR, E5-94-189, Dubna, 1994}\\[5ex]

\noindent http://www.arXiv.org e-Print archive No.~math-ph/0303038\\[3ex]

\noindent 2003 PACS numbers: 02.40Vh, 04.90.+e

\noindent
{\small
The \LaTeXe\ source file of this paper was produced by converting a
ChiWriter 3.16 source file into
ChiWriter 4.0 file and then converting the latter file into a
\LaTeX\ 2.09 source file, which was manually edited for correcting numerous
errors and for improving the appearance of the text.  As a result of this
procedure, some errors in the text may exist.
}\\[7ex]

 {\bf 1. INTRODUCTION}
\nopagebreak

\medskip
 The work deals with the derivation of first order (with respect to a
corresponding parameter) equations satisfied by the introduced in [1]
relative mechanical quantities in spaces (manifolds) whose tangent bundle is
endowed with a transport along paths, which in the present paper is supposed
to be linear [2]. These equations, in fact, were found in [3] in a component
form, i.e., in a fixed local (and coordinate) basis. Besides, in [3]
implicitly were used linear transports along paths without
self-intersections. The present investigation closely follows the ideas of
[3], but to make a difference of it, here are used a coordinate independent
language (where it is possible) and linear transports along arbitrary paths.

As the mentioned equations have a form and physical interpretation similar to that of the equation of geodesic deviation (of first order for the infinitesimal deviation vector) [4,5], they are called deviation equations (for the corresponding quantities).

Section 2 contains certain approximate results concerning linear transports along paths in vector bundles. In sections 3 through 7 the first order deviation equations are derived for, respectively, the deviation vector, relative velocity, relative momentum, relative acceleration and the relative energy of two arbitrary moving point particles. Also connections between these quantities are found.

Below, for reference purposes, some definitions and constructions are presented.

All considerations in the present work, with an exception of a
 part of Sect. 2, are made in a (real) differentiable manifold $M [6,7]$
whose tangent bundle $(T(M),\pi ,M)$ is endowed with a linear transport
$(L$-transport) along paths [2] and a covariant differentiation (linear
connection$) \nabla  [6,7]$. Here $T(M):=\cup_{x\in M}T_{x}(M), T_{x}(M)$
being the tangent to $M$ space at $x\in M$ and $\pi :T(M)  \to M$ is such
that $\pi (V):=x$ for $V\in T_{x}(M)$.

By $J$ and $\gamma :J  \to M$ we denote, respectively, an arbitrary real interval and a path in M. If $\gamma $ is of class $C^{1}$, its tangent vector is written as   .

The linear transport $(L$-transport) along paths in $(T(M),\pi ,M) (cf. [2])$
is a map $L:\gamma  \to L^{\gamma }, L^{\gamma }:(s,t)\to L^{\gamma }_{s  \to
t}, s,t\in J$ being the $L$-transport along $\gamma $, where $L^{\gamma }_{s
\to t}:T_{\gamma (s)}(M)  \to T_{\gamma (t)}(M)$, satisfy the equalities
\[
  L^{\gamma }_{t  \to r}\circ L^{\gamma }_{s  \to t}=L^{\gamma }_{s  \to r},
\quad r,s,t\in J,\qquad (1.1)
\]
\[
L^{\gamma }_{s  \to s}={\it id}_{T_{\gamma (s)}}, \qquad  s\in J,\qquad
(1.2)
\]
\[
  L^{\gamma }_{s  \to t}(\lambda U+\mu V)=\lambda L^{\gamma }_{s  \to t}U+\mu
L^{\gamma }_{s  \to t}V, \qquad  s,t\in J, U,V\in T_{\gamma (s)}(M).
 \qquad (1.3)
\]
 Here {\it id}$_{X}$is the identity map of the set X.

If $X, Y$, and $Z$ are vector fields on $M [7]$, then the tensors
(operators) of torsion $T$ and curvature $R$ of the covariant differentiation
$\nabla $, respectively, are [7]
\[
 T(X,Y):=\nabla _{X}Y-\nabla _{Y}X-[X,Y],\qquad (1.4)
\]
\[
  R(X,Y)Z:=\nabla _{X}\nabla _{Y}Z-\nabla _{Y}\nabla _{X}Z-\nabla
_{[X,Y]}Z,\qquad (1.5)
\]
 where $[X,Y]$ is the commutator of $X$ and Y.

The covariant differentiation (derivative) along the $C^{1}$path $\gamma :J
\to M$, i.e. $\nabla _{\dot\gamma }$, will be denoted by $D/ds\mid _{\gamma
}, s\in J$, or simply by $D/ds$ if there is no risk of misunderstanding.

Let there be given paths $x_{a}:J  \to M, a=1,2$ and a one-parameter family
of paths $\{\gamma _{s}:J^\prime   \to M, s\in J\}$ such that $\gamma
_{s}(r^\prime ):=x_{1}(s)$ and $\gamma _{s}(r^{\prime\prime}):=x_{2}(s)$ for
some $r^\prime ,r^{\prime\prime}\in J^\prime $. The tangent vectors to the
paths $\gamma _{s}:r\to \gamma _{s}(r)$ and $\gamma _{.}(r):s\to \gamma
_{s}(r), s\in J, r\in J^\prime $ are denoted, respectively by
$\dot\gamma_{s}$ and $\gamma {^\prime }(r)$.

The differentiation $D/ds\mid _{x_{1}}$will for, brevity, be written as
$D/ds$.

 The deviation vector of $x_{2}$with respect to $x_{1}$at $x_{1}(s) (cf. [8],
eq. (2.5))$ is
\[
 h_{21}=h_{21}\mid _{\gamma _{s}}
=h(s;x_{1})
= \int^{r^{\prime\prime}}_{r^\prime }
(L^{\gamma _{s}}_{u  \to r^\prime} \dot\gamma_{s}(u))du.  \qquad (1.6)
\]

 Let the paths $x_{1}$ and $x_{2}$ be world lines of the point particles,
respectively, 1 and 2. Their velocities [4], the relative velocity (of 2 with
respect to $1; cf. [1])$, and the corresponding to them accelerations,
respectively, are:
\[
 V_{1}=\dot{x}_{1},\quad V_{2}=\dot{x}_{2},\qquad (1.7a)
\]
\[
  \Delta V_{21}= L^{\gamma _{s}}_{r^{\prime\prime}  \to r^\prime }V_{2}-
V_{1},\qquad (1.7b)
\]
\[
 A_{a}=\frac{D}{ds}\big|_{x_{1}}V_{a},\quad a=1,2,\qquad (1.8a)
\]
\[
  \Delta A_{21}= L^{\gamma _{s}}_{r^{\prime\prime}  \to r^\prime }A_{2}-
A_{1}.\qquad (1.8b)
\]

 The momenta of the considered particles are [4,1]
\[
 p_{a}=\mu _{a}V_{a}, \qquad  a=1,2,
\]
  where $\mu _{a}:J  \to {\Bbb R}\backslash \{0\}, a=1,2$ are (nonvanishing)
scalar functions (identified with the corresponding proper masses if the
latter are nonzero; $cf.[4])$.

 The relative momentum of the second particle with respect to the first one
is $(cf. [1]$, sect. 3)
 \[
  \Delta p_{21}= L^{\gamma _{s}}_{r^{\prime\prime}  \to r^\prime }p_{2}-
p_{1}.\qquad (1.9)
\]

\medskip
\medskip
 {\bf 2. SOME APPROXIMATE RESULTS FOR\\ LINEAR TRANSPORTS ALONG PATHS}

\medskip
 In this section, approximate results concerning linear transports
$(L$-transports) along paths in vector bundles will be obtained. They will be
used later on in the present work. For details of the theory of
$L$-transports the reader is referred to [2], only a few facts of which will
be cited below.

 Let $L^{\gamma }$ be an $L$-transport along the path $\gamma :J  \to B$ in
the vector bundle $(E,\pi ,B) ($see $[2]; cf$. Sect. 1) and $\{e_{i}(s):
i=1,\ldots  ,\dim(\pi ^{-1}(x)), x\in B\}$ be a bases in $\pi ^{-1}(\gamma
(s)), s\in $J. Let in the basis $\{e_{i}(r)\}, r=s,t$ the transport
$L^{\gamma }_{s  \to t}$along $\gamma $ from $s$ to $t, s,t\in J$ be
described by the matrices $H(t,s;\gamma ):=  H^{i}_{.j}(t,s;\gamma )  $. So
$(cf. [2]$, sect. $2, eq. (2.10))$ if $u_{s}=u^{i}_{s}e_{i}(s)\in \pi
^{-1}(\gamma (s)), s\in J$, where here and hereafter the Latin indices run
from 1 to $\dim(\pi ^{-1}(x)), x\in B$ and summation from 1 to $\dim(\pi
^{-1}(x)), x\in B$ over repeated on different levels indices is assumed. Then
\[
 L^{\gamma }_{s  \to t}u_{s}=H^{i}_{.j}(t,s;\gamma
)u^{j}_{s}e_{i}(t).\qquad (2.1)
\]

 If $\gamma $ and $H$ are $C^{N+1}$functions, $N$ being an integer, then due
to $H(s,s;\gamma )={\Bbb I}$, with ${\Bbb I}$ the unit matrix, (see $[2], eq.
(2.12))$, the expansions:
 \[
  H(t,s;\gamma )
={\Bbb I} + \sum_{m=1}^N  \frac{1}{m!}
H_{i_{1}\ldots i_m}(s;\gamma)
(\gamma ^{i_{1}}(t)-\gamma ^{i_{1}}(s))\cdot \cdot \cdot (\gamma
^{i_{m}}(t)-\gamma ^{i_{m}}(s))
\]
\[
+O((t-s)^{N+1}),\qquad (2.2a)
\]
\[
 H(t,s;\gamma )
={\Bbb I}+ \sum_{m=1}^N  \frac{1}{m!}
^{m}H(s;\gamma )(t-s)^{m}+O((t-s)^{N+1})\qquad (2.2b)
\]
 are valid, where
\[
  H_{i_{1}\ldots i_m}(s;\gamma )
:= \frac{\partial^m H(t,s;\gamma)}
	{\partial\gamma^{i_m}\ldots\partial\gamma^{i_1}}
\Big|_{t=s},\qquad (2.3a)
\]
\[
^{m}H(s;\gamma ):=
\frac{\partial^m H(t,s;\gamma)}{\partial t^m}
\Big|_{t=s}.\qquad (2.3b)
\]

 The matrices (2.3b) can easily be expressed through the matrices (2.3a),
e.g. we have:
\[
^{1}H(s;\gamma )=H_{i}(s;\gamma )\dot\gamma^{i}(s),\qquad (2.4a)
\]
\[
 ^{2}H(s;\gamma )
=H_{ij}(s;\gamma ) \dot\gamma^{i}(s)\dot\gamma^{j}(s)
+ H_{i}(s;\gamma ) \frac{d\dot\gamma(s)}{ds} . \qquad (2.4b)
\]

 If in (2.1) we substitute $H(t,s;\gamma )$ with its $N$-th approximation
with respect to $t-s$, which due to (2.2) is
\[
 ^{(N)}H(t,s;\gamma )
={\Bbb I} + \sum_{m=1}^N  \frac{1}{m!}
H_{i_{1}\ldots i_m}(s;\gamma )(\gamma ^{i_{1}}(t)-\gamma ^{i_{1}}(s))
\cdot \cdot \cdot
(\gamma ^{i_{m}}(t)-\gamma ^{i_{m}}(s))
\]
\[
 ={\Bbb I}+ \sum_{m=1}^N  \frac{1}{m!}
^m H(s;\gamma )(t-s)^{m},\qquad (2.5)
\]
 where the second equality is with a precision of $O((t-s)^{N+1})$, we get a
map $^{(N)}L^{\gamma }_{s  \to t}:\pi ^{-1}(\gamma (s))  \to \pi ^{-1}(\gamma
(t))$ defined in $\{e_{i}\}$ by
\[
 ^{(N)}L^{\gamma }_{s  \to t}u_{s}:=^{(N)}H^{i}_{.j}(t,s;\gamma
)u^{j}_{s}e_{i}(s).\qquad (2.6)
\]

Evidently, $^{(N)}L^{\gamma }$is the $N$-th approximation to $L^{\gamma }($in $\{e_{i}\})$, i.e., they coincide up to terms of $(N+1)$-th order with respect to the difference $t$-s.

Further in this paper we shall work only with the zero-th $(N=0)$ and first
$(N=1)$ approximations to $L^{\gamma }$, which according to (2.5) are
described by
\[
 ^{(0)}H(t,s;\gamma )={\Bbb I},\qquad (2.7a)
\]
\[
 ^{(1)}H(t,s;\gamma )={\Bbb I}+H_{i}(s;\gamma )(\gamma ^{i}(t)-\gamma
^{i}(s))
\]
\[
 ={\Bbb I}+^{1}H(s;\gamma )(t-s)={\Bbb I}+H_{i}(s;\gamma
)\dot\gamma^{i}(s)(t-s).\qquad (2.7b)
\]
(The second and third equalities in (2.7b) are up to $O((t-s)^{2}).)$

Let us note that the zero-th approximation (2.7a) is constant and, hence, it depends neither on $s, t$ and $\gamma $, nor on the used $L$-transport.

From here on in this work we will consider only the case of the tangent bundle to a differentiable manifold $M$, i.e. it is supposed that $(E,\pi ,B)=(T(M),\pi ,M)$.

It is important to be emphasized that in the case $(E,\pi ,B)=(T(M),\pi ,M)$ the components of the involved in (2.7b) matrices $(-H_{i}(s;\gamma ))$, i.e. $[-H^{i}_{.jk}(s;\gamma )]=-[\partial H^{i}_{.j}(t,s;\gamma )/\partial \gamma ^{k}(t)]\mid _{t=s}$, are coefficients of an affine connection along $\gamma $, i.e. under a change of the basis $e_{i}(s)$ they transform like usual coefficients of an affine connection [10]. This proposition is a simple corollary of (2.3a) for $m=1$ and the circumstance that $H^{i}_{.j}(s;\gamma )$ in this case are components of a two-point tensor from $T_{\gamma (t)}\otimes T^{*}_{\gamma (s)}($see [2], sect. 2). We shall note here, without a proof, that the so arising connection with the coefficients $[-H^{i}_{.jk}(s;\gamma )]$ along $\gamma $ is flat, i.e., its curvature tensor is zero.

If in the tangent bundle $(T(M),\pi ,M)$ an affine connection with the
coefficients $\{\Gamma ^{i}_{.jk}(x)\} ($see e.g. [7]) is given, then it is
easy to calculate that for the parallel transport defined by the connection
\[
 H^{i}_{.jk}(s;\gamma ; || )=-\Gamma ^{i}_{.jk}(\gamma (s))\qquad (2.8)
\]
is valid, where an additional argument  $||$   indicates that the
calculations are made for the pointed parallel transport. According to the
above said, this equality means that connection along any path $\gamma $
induced by the parallel transport coincides with the restriction of the
affine connection on the path $\gamma $ generating this transport.

Let there be given paths $x_{a}:J  \to M, a=1,2$ and a one-parameter family of paths $\{\gamma _{s}:J^\prime   \to M, s\in J\}$ such that $\gamma _{s}(r^\prime ):=x_{1}(s)$ and $\gamma _{s}(r^{\prime\prime}):=x_{2}(s)$ for some $r^\prime ,r^{\prime\prime}\in J^\prime $.

Let in $M$ be given an affine connection $\nabla $ with local coefficients
$\Gamma ^{i}_{.jk}[7], B$ be $a C^{1}$ vector field on $\{\gamma _{s}(r):
r\in J^\prime , s\in J\}$ and
\[
  \Delta B_{21}:=L^{\gamma _{s}}_{r^{\prime\prime}  \to r^\prime
}B_{x_{2}}-B_{x_{1}}\in T_{x_{1}}(M).\qquad (2.9)
\]

 Applying to the first term of this definition (2.1) and (2.2b) for $N=1$ and
taking into account $(2.4a), x_{1}(s)=\gamma _{s}(r^\prime )$ and
$x_{2}(s)=\gamma _{s}(r^{\prime\prime})$, we find after some simple
calculations
\[
\Delta B_{21}
= \Bigl(\frac{D^H}{dr}\Big|_{\gamma_s} B\Bigr)
\Big|_{\gamma_{s}(r')} (r^{\prime\prime}-r^\prime )
+ O((r^{\prime\prime}-r^\prime )^{2})
=
= \Bigl(\frac{D}{dr}\Big|_{\gamma_s} B\Bigr)
\Big|_{\gamma_{s}(r')} (r^{\prime\prime}-r^\prime )
\]
\[
+ S(B,\zeta_2)|_{\gamma_s(r')} +
+ O((r^{\prime\prime}-r^\prime )^{2})
\]
where $D/dr\mid _{\gamma _{s}}:=\nabla _{\dot\gamma_{s}}$ is the covariant
derivative along $\gamma _{s}:J^\prime   \to M$ generated by $\nabla ,
D^{H}/dr\mid _{\gamma _{s}}:=\nabla _{\dot\gamma_{s}}\big|_{\Gamma
^{i}_{.jk}}$ is the covariant derivative along $\gamma _{s}$generated by the
connection with local coefficients $-H^{i}_{.jk}$and $S$ is a tensor field of
the type (1,2) whose components in any local basis are
\[
 S^{i}_{.jk}|_{\gamma _{s}}:=-H^{i}_{.jk}(r;\gamma _{s})-\Gamma
^{i}_{.jk}(\gamma _{s}(r)), \qquad  r\in J^{\prime\prime},\ s\in J,\qquad
(2.11)
\]
i.e. $(S(B,\zeta _{21})\mid _{\gamma (r^\prime )})^{i}=S^{i}_{.jk}$
$_{\gamma (r^\prime )}B^{j}_{\gamma _{s}}\zeta ^{k}_{21}|_{\gamma
(r^\prime )}$. Here, as in (2.10),
\[
 \zeta _{21}\mid _{\gamma _{s}}
=(r^{\prime\prime}-r^\prime )\dot\gamma_{s}(r^\prime )\qquad (2.12)
\]
 is the infinitesimal deviation vector at $\gamma _{s}(r^\prime ) (cf. [8],
eq.  (2.10))$ which due to (2.7b) (see also $[8], eq. (2.9))$ is connected
with the deviation vector (1.6) of $x_{2}$with respect to $x_{1}$through the
equality
\[
  h_{21}=\zeta _{21}+O((r^{\prime\prime}-r^\prime )^{2}).\qquad (2.13)
\]

In this equality, as well as till the end of this work, we will work only with a precision of terms up to $O((r^{\prime\prime}-r^\prime )^{2})$.

\medskip
\medskip
 {\bf 3. EQUATION FOR THE DEVIATION VECTOR}
\nopagebreak

\medskip
In fact, the equation, mentioned in the title of this section, has been
derived in [8], example 3.2 and it is expressed by equation (3.27) of [8]
which, in our case equivalently, can be written as
\[
\frac{D^2 \zeta_{21}}{ds}
= R(V_{1},\zeta _{21})V_{1}+ T(V_{1},D\zeta _{21}/ds) +
\frac{DT}{ds} (V_{1},\zeta _{21}) + T(F_{s},\zeta _{21})
\]
\[
+
\frac{DF_s(r)}{dr} \Big|_{r=r^\prime } (r^{\prime\prime}-r^\prime ) +
O((r^{\prime\prime}-r^\prime )^{2}),\qquad (3.1)
\]
 where $V_{1}$is the velocity of the first particle, the force field
$F_{s}$is defined by
\[
 F_{s}(r):=\nabla _{\gamma ^\prime }\gamma ^\prime|_{\gamma _{s}}
={\bigl(} \frac{D}{da}\Big|_{\gamma _{\cdot }(r)} \gamma ^\prime
{\bigr)}_{\gamma _{s}}\qquad (3.2)
\]

 and all quantities are taken at the point $\gamma _{s}(r^\prime )$.

We want to emphasize two features of equation (3.1). Firstly, it is independent of the concrete choice of the used transport along paths $L$, and also (up to terms $O((r^{\prime\prime}-r^\prime )^{2})))$ of the family $\{\gamma _{s}\}$. And secondly, the derivation of this equation in [3] shows that in it the correction $O((r^{\prime\prime}-r^\prime )^{2})$ is strictly equal to zero.

$Eq. (3.1)$ is a generalization, in the case of arbitrary paths
in spaces with torsion, of the classical equation of geodesic deviation [4,5].

\medskip
\medskip
 {\bf 4. EQUATION FOR THE RELATIVE VELOCITY}
\nopagebreak

\medskip
If along $x_{1}$and $x_{2}$particles 1 and 2 are moving, respectively, then
their velocities are given by (1.7a). Fist of all we want to find the
connection between their relative velocity (1.7b) and the first order
deviation velocity $D\zeta _{21}/ds$ which due to (2.13) is connected with
the deviation velocity $Dh_{21}/ds$ through the equality
 \[
  Dh_{21}/ds=D\zeta _{21}/ds+O((r^{\prime\prime}-r^\prime )^{2}).\qquad (4.1)
\]

 Taking into account  $x_{1}(s)=\gamma _{s}(r^\prime )$  and
$x_{2}(s)=\gamma _{s}(r^{\prime\prime})$,  we  find $\zeta
^{i}_{21}(s)=x^{i}_{2}(s)-x^{i}_{1}(s)+O((r^{\prime\prime}-r^\prime )^{2})$
and
\(
V^{i}_{2}(s)-V^{i}_{1}(s)
=\frac{\partial^2 \gamma^i_s(r)}{\partial r\partial s}
\Big|_{r=r^\prime }  (r^{\prime\prime}-r^\prime)
 +O((r^{\prime\prime}-r^\prime )^{2}).
\)
 Differentiating covariantly the
former of these equa  tions with respect to $s$ along $x_{1}$and using the
latter one, we get
\[
\frac{D\zeta_{21}}{ds}
= \Bigl(\frac{D}{Ds}\Big|_{\gamma_s} \gamma'_s\Bigr)\Big|_{r=r'}
(r^{\prime\prime}-r^\prime )+T(V_{1},\zeta _{21})|_{x_{1}}
+O((r^{\prime\prime}-r^\prime )^{2}).
 \qquad (4.2)
\]

 On the other hand, from (2.10) for $B_{\gamma _{s}}=\gamma $  $(r), (2.9)$,
and (1.7) we see that
\[
\Delta V_{21}
=  \Bigl(\frac{D}{Ds}\Big|_{\gamma_s} \gamma'_s\Bigr)\Big|_{r=r'}
 (r^{\prime\prime}-r^\prime )+S(V_{1},\zeta _{21})|_{x_{1}}
+O((r^{\prime\prime}-r^\prime )^{2}). (4.3)
\]
 Comparing this equality with (4.2), we find the following relation between
the relative velocity $\Delta V_{21}$and the first order deviation velocity
$D\zeta _{21}/ds$:
\[
\frac{D\zeta_{21}}{ds}
=\Delta V_{21}+{\bigl(}T(V_{1},\zeta _{21})-S(V_{1},\zeta _{21})
\bigr)|_{x_{1}}+O((r^{\prime\prime}-r^\prime )^{2}).\qquad (4.4)
\]

 From here we can make the conclusion that up to second order terms the
deviation velocity describes the "general relative velocity" of particle 2
with respect to particle 1. It is caused by the (nongravitational)
interaction between the particles and all properties of the manifold $M
($curvature, torsion, transport along paths). The relative velocity (1.7b) is
caused by the (nongravitational) interaction of the particles and the
transport along paths used.

Sub stituting (4.4) into the left-hand side of (3.1) and performing some
evident transformations, we get the {\it deviation equation for the relative
velocity} $\Delta V_{21}$in the form
\[
\frac{DV_{21}}{ds}
=R(V_{1},\zeta _{21})V_{1}
+ \frac{D}{ds} [S(V_{1},\zeta _{21})]
+ \frac{DF_s(r)}{dr}\Big|_{r=r^\prime }
(r^{\prime\prime}-r^\prime)
+O((r^{\prime\prime}-r^\prime )^{2}),  \qquad (4.5)
\]
 where all quantities are evaluated at $x_{1}(s)$.

This equation describes up to second order terms the change of the relative velocity of the second particle with respect to the first one along the world line of the latter.

\medskip
\medskip
 {\bf 5. EQUATION FOR THE RELATIVE MOMENTUM}
\nopagebreak

\medskip
In our case (see Sect. 1) due to $[1], eq. (3.6)$, the relative momentum of
the particles $\Delta p_{21}$defined by (1.9), is
\[
 \Delta p_{21}=\mu _{2}(s)\Delta V_{21}+[\mu _{2}(s)/\mu
_{1}(s)-1]p_{1},\qquad (5.1)
\]
 which may be obtained also as an approximate result from $(2.9), (4.3)$ and
(2.10) for $B_{\gamma _{s}}=\mu (s,r)\gamma $  (r) with $\mu $ being $a
C^{1}$function such that $\mu (s,r^\prime )=\mu _{1}(s)$ and $\mu
(s,r^{\prime\prime})=\mu _{2}(s)$. Differentiating this equality covariantly
along $x_{1}$, we get
\[
\frac{D\delta p_{21}}{ds}
=\mu _{2}(s)\frac{D\delta V_{21}}{ds}
+ \frac{D\mu_2(s)}{ds} \Delta V_{21}
+ \frac{D}{ds}[(\mu _{2}(s)/\mu _{1}(s)- 1)p_{1}(s)]
\]
 and substituting here (4.5), we obtain
\[
\frac{D\delta p_{21}}{ds}
= \frac{\mu_2(s)}{(\mu_1(s))^2 }
 R(p_{1},\zeta _{21})p_{1}
 +  \mu _{2}(s) \frac{D}{ds} \Bigl[ \frac{1}{\mu_1(s)}S(p_{1},\zeta_{21})
\Bigr]
+ \frac{D\mu_2(s)}{ds} \Delta V_{21}+
\]
\[
+\frac{D}{ds}
\Bigl[ \Bigl( \frac{\mu_2(s)}{\mu_1(s)} - 1\Bigr) p_1(s) \Bigr]
+ \mu_2(s) \frac{DF_s(r)}{dr}\Big|_{r=r'}
(r^{\prime\prime}-r^\prime )+O((r^{\prime\prime}-r^\prime )^{2}).
 \qquad (5.2)
\]
  Here, if it is necessary, $\Delta V_{21}$may be substituted with the
obtained for it expressions from (5.1) or (4.3).

This is the {\it first order deviation equation for the relative momentum} of the second particle with respect to the first one. It describes the evolution of $\Delta p_{21}$along the trajectory of the first particle. The physical interpretation of (5.2) will be considered below in section 7.

\medskip
\medskip
 {\bf 6. EQUATION FOR THE RELATIVE ACCELERATION}
\nopagebreak

\medskip
From the definitions (1.8a) and (3.2) we find the following representation
for the accelerations of the particles studied:
\[
 A_{1}=F_{s}(r^\prime ),
 \qquad  A_{2}=F_{s}(r^{\prime\prime}).\qquad (6.1)
\]
 This is very natural as from (3.2) and the physical interpretation of the
deviation equation (3.1) (see [8]) it is clear that $F_{s}(r)$ has a sense of
a (nongravitational) force per unit mass acting on a particle situated at the
point $\gamma _{s}(r)$; i.e., $F_{s}(r)$ is the acceleration of that
particle.

If in (2.9) and (2.10) we let $B_{\gamma _{s}}=F_{s}(r)$, we find the
relative acceleration (1.8b) of the particles in the form
\[
 \Delta A_{21}
=\Delta F_{21}
= \frac{D}{dr}\Big|_{\gamma _{s}}F_{s}(r)|_{r=r^\prime }
(r^{\prime\prime}-r^\prime )+S(A_{1},\zeta _{21})|_{x_{1}}
+O((r^{\prime\prime}-r^\prime )^{2}).  \qquad (6.2)
\]

The first order deviation acceleration between the considered particles is
$D^{2}\zeta _{21}/ds^{2}$and according to (2.13) is connected with the
deviation acceleration $D^{2}h_{21}/ds^{2}$by
\[
 D^{2}h_{21}/ds^{2}=D^{2}\zeta _{21}/ds^{2}+O((r^{\prime\prime}-r^\prime
)^{2}).\qquad (6.3)
\]

Expressing $DF_{s}(r)/ds\mid _{r=r^\prime }(r^{\prime\prime}-r^\prime )$
from (6.2) and substituting the so-obtained result into (3.1), we get the
following relation between the first order deviation acceleration $D^{2}\zeta
_{21}/ds^{2}$and the relative acceleration $\Delta A_{21}$:
\[
\frac{D^2\zeta_{21}}{ds^2}
= \Delta A_{21}+R(V_{1},\zeta _{21})V_{1}+T(A_{1},\zeta _{21})-S(A_{1},\zeta
_{21})+T(V_{1},D\zeta _{21}/ds)
\]
\[
+ \frac{DT}{ds}(V_{1},\zeta _{21})
+O((r^{\prime\prime}-r^\prime )^{2}).\qquad (6.4)
\]
  Here, for some purposes, it is convenient to replaced the first order
deviation velocity $D\zeta _{21}/ds$ with the right-hand side of (4.4).

The last equation shows that up to second order terms the deviation acceleration is caused by the (nongravitational) interaction between the particles and the properties of the space M. The cause for the relative acceleration is only the (nongravitational) interaction between the particles and the $L$-transport along paths used.

If we express $DF_{s}(r)/ds\mid _{r=r^\prime }(r^{\prime\prime}-r^\prime )$
from (4.5) and substitute the result into (6.2), we shall find the following
relation between the relative velocity $\Delta V_{21}$and the relative
acceleration $\Delta A_{21}$:
\[
 \Delta A_{21}=
\frac{D\Delta V_{21}}{ds}
- R(V_{1},\zeta _{21})V_{1}
+ S(V_{1},\frac{\zeta_{21}}{ds}) +
 \frac{DS}{ds}(V_{1},\zeta _{21})
+ O((r^{\prime\prime}-r^\prime )^{2}).  \qquad (6.5)
\]

In accordance with the physical interpretation of the involved in this equation quantities it can be called a {\it first order deviation equation for the relative acceleration.}

\medskip
\medskip
 {\bf 7. EQUATION FOR THE RELATIVE ENERGY}
\nopagebreak

\medskip
Before the derivation of the equation mentioned in the title, we will write
the deviation equation for the relative momentum (5.2) in a form which is a
direct analog of the second Newton's law, i.e., as an equation of motion for
the considered case.

Let $\mu :J  J  \to {\Bbb R}$ be $a C^{1}$function, $\mu (s,r^\prime )=\mu
_{1}(s), \mu (s,r^{\prime\prime})=\mu _{2}(s)$ and $K(s,r):=\mu (s,r)F_{s}(r)
($see (3.2)). The quantity $K(s,r)$ has a meaning of a (nongravitational)
force acting on a particle with momentum $\mu (s,r)\gamma $  (r) situated at
$\gamma _{s}(r)$. Putting $B_{\gamma _{s}}=K(s,r)$ into (2.10) and taking
into account (6.1) and
\[
 \mu _{2}(s)-\mu _{1}(s)=\mu (s,r^{\prime\prime})-\mu (s,r^\prime )
= \frac{\partial\mu(s,r)}{\partial r} \Big|_{r=r^\prime }
(r^{\prime\prime}-r^\prime )+O((r^{\prime\prime}-r^\prime )^{2}),
\]
we get
\[
\Delta K_{21}
=(\mu _{2}(s)-\mu _{1}(s))A_{1}
+ \mu _{1}(s) \frac{DF_s(r)}{dr}\Big|_{r=r^\prime }
(r^{\prime\prime}-r^\prime )+\mu _{1}(s)S(A_{1},\zeta _{21})
\]
\[
+O((r^{\prime\prime}-r^\prime )^{2})
= (\mu _{2}(s)-\mu _{1}(s))A_{1}
+ \mu _{2}(s) \frac{DF_s(r)}{dr}\Big|_{r=r^\prime }
(r^{\prime\prime}-r^\prime )
\]
\[
 +\mu _{2}(s)S(A_{1},\zeta _{21})+O((r^{\prime\prime}-r^\prime )^{2}).\qquad
(7.1)
\]

Physically $\Delta K_{21}$is the (covariant) difference between the forces
acting on the particles studed.

Expressing the term $\mu _{2}(s)DF_{s}(r)/dr\mid _{r=r^\prime
}(r^{\prime\prime}-r^\prime )$ from (7.1) and substituting the result into
(5.2), we obtain
\[
\frac{Dp_{21}}{ds}
= \frac{\mu_2(s)}{(\mu_1(s))^2} R(p_{1},\zeta _{21})p_{1}
+ \frac{\mu_2(s)}{\mu_1(s)}
\Bigl[ S(p_{1},D\zeta _{21}/ds)+ \frac{DS}{ds}(p_{1},\zeta_{21})
\Bigr]
\]
\[
+ \frac{d\mu_2(s)}{ds} \Delta V_{21}
+ \frac{1}{\mu_1(s)}
\frac{d(\mu_2(s)-\mu_1(s))}{ds} p_{1}
+ \Delta K_{21}+O((r^{\prime\prime}-r^\prime )^{2}).   \qquad (7.2)
\]

This is the first order deviation equation for the relative momentum in the form of an equation of motion.

Let in the tangent bundle $(T(M),\pi ,M)$ be given also a real bundle metric $g ($besides the transport along paths I and the covariant differentiation $\nabla )$, i.e., $[6] a$ map $g:x\to g_{x}, x\in M$, where the maps $g_{x}:T_{x}(M)\otimes T_{x}(M)  \to {\Bbb R}$ are bilinear, nondegenerate and symmetric. For brevity, the scalar products of $X,Y\in T_{y}(M), y\in M$ defined by $g$ will be denoted by a dot $(\cdot )$, i.e. $X\cdot Y:=g_{y}(X,Y)$. The scalar square of $X$ will be written as $(X)^{2}$for it has to be distinguished from the second component $X^{2}$of $X$ in some local basis (in a case when $\dim(M)>1)$. As $g$ is not supposed to be positively defined, $(X)^{2}$can take any real values.

Then, according to $[1], eq. (4.1)$, in the considered case the relative
energy of the second particle with respect to the first one is
\[
 E_{21}=\epsilon ((V_{1}(s))^{2})(I^{\gamma _{s}}_{r^{\prime\prime}  \to
r^\prime }p_{2}(s))\cdot V_{1}(s)
\]
\[
=\epsilon ((V_{1}(s))^{2})(\Delta p_{21}\cdot V_{1}+p_{1}\cdot V_{1}),\qquad
(7.3)
\]
where we have used (1.9).

Differentiating (7.3) with respect to $s$ along $x_{1}$and substituting the
obtained result into (7.2), we find:
\[
\frac{dE_{21}}{ds}
= \epsilon ((V_{1})^{2})
\Big\{
  \frac{\mu_2(s)}{(\mu_1(s))^3}[R(p_{1},\zeta _{21})p_{1}]\cdot p_{1}
+ \frac{\mu_2(s)}{(\mu_1(s))^2} p_{1}\cdot
\Big[
S(p_{1},\frac{D\zeta_{21}}{ds})
\]
\[
+ \frac{DS}{ds}(p_{1},\zeta _{21})\Big]
+ \frac{d\mu_2(s)}{ds} V_{1}\cdot \Delta V_{21}
+ \frac{1}{(\mu_1(s))^2}
  \frac{d(\mu_2(s)-\mu_1(s))}{ds} p_{1}\cdot p_{1}
+ V_{1}\cdot \Delta K_{21}
\]
\[
+ \frac{\Delta g}{ds} (\Delta p_{21},V_{1})
+ \Delta p_{21}\cdot A_{1}
+ \frac{d\mu_1(s)}{ds} V_{1}\cdot V_{1}
+ \mu _{1}(s)(2V_{1}\cdot A_{1}
+ \frac{Dg}{ds}(V_{1},V_{1}))
\]
\[
+ O((r^{\prime\prime}-r^\prime )^{2}),\qquad (7.4)
\]
where all quantities are taken at the point $x_{1}(s)$.

This is the {\it first order deviation equation for the relative energy}. It has a meaning of an equation for the energy balance and
can be considered as a generalization of the energy conservation equation in the situation studied.

\medskip
\medskip
 {\bf ACKNOWLEDGEMENT}

\medskip
This research was partially supported by the Fund for Scientific Research of Bulgaria under contract Grant No. $F 103$.

\medskip
\medskip
 {\bf REFERENCES}

\medskip
1.  Iliev B.Z., Relative mechanical quantities in spaces with a transport along paths, JINR Communication $E2-94-188$, Dubna, 1994.\par
2.  Iliev B.Z., Linear transports along paths in vector bundles. I. General theory, Communication JINR, $E5-93-239$, Dubna, 1993.\par
3.  Iliev B.Z., On the first order deviation equations and some quantities connected with them, Communication JINR, $E2-90-534$, Dubna, 1990.\par
4.  Synge J.L., Relativity: The general theory, North-Holland Publ. Co., Amsterdam, 1960.\par
5.  Weber J., General Relativity and Gravitational Waves, New York, 1961.\par
6.  Dubrovin B.A., S.P. Novikov, A.T. Fomenko, Modern geometry, Nauka, Moscow, 1979 (In Russian).\par
7.  Kobayashi S., K. Nomizu, Foundations of differential geometry, vol.1, Interscience publishers, New-York-London, 1963.\par
8.  Iliev B.Z., Deviation equations in spaces with a transport along paths, JINR Communication $E2-94-40$, Dubna, 1994.

\newpage

\medskip
\medskip
 Iliev B. Z.\\[5ex]

\noindent
 Firs Order Deviation Equations\\
 in Spaces with a Transport along Paths\\[5ex]

\medskip
\medskip
In a coordinate free form are found the (deviation) equations satisfied by
the (infinitesimal) deviation vector, relative velocity, relative momentum,
relative acceleration and relative energy of two point particles in a
differentiable manifold the tangent bundle of which is endowed with a linear
transport along paths, a linear connection and, in the last case, also with a
metric. Some approximate relations between these quantities are
obtained.\\[5ex]

\medskip
\medskip
 The investigation has been performed at the Bogoliubov Laboratory of
Theoretical Physics, JINR.

\end{document}